\documentclass[aps,prc,showpacs,twocolumn,groupedaddress]{revtex4}

\usepackage{graphics}

\begin{document}


\title{Phases of hot nuclear matter at subnuclear densities}


\author{Gentaro Watanabe$^{a,b,c}$, Katsuhiko Sato$^{a,d}$,
  Kenji Yasuoka$^{e}$ and Toshikazu Ebisuzaki$^{b}$}
\affiliation{
$^{a}$Department of Physics, University of Tokyo,
Tokyo 113-0033, Japan
\\
$^{b}$Computational Astrophysics Laboratory, RIKEN,
Saitama 351-0198, Japan
\\
$^{c}$NORDITA, Blegdamsvej 17, DK-2100 Copenhagen \O, Denmark
\\
$^{d}$Research Center for the Early Universe, 
University of Tokyo,
Tokyo 113-0033, Japan
\\
$^{e}$Department of Mechanical Engineering, Keio University,
Yokohama 223-8522, Japan}


\date{\today}

\begin{abstract}
Structure of hot dense matter
at subnuclear densities is investigated
by quantum molecular dynamics (QMD) simulations.
We analyze nucleon distributions and nuclear shapes using
two-point correlation functions and Minkowski functionals
to determine the phase-separation line
and to classify the phase of nuclear matter
in terms of the nuclear structure.
Obtained phase diagrams show that the density of
the phase boundaries between the different nuclear structures
decreases with increasing temperature due to the thermal expansion
of nuclear matter region.
The critical temperature for the phase separation
is $\agt 6$ MeV for the proton fraction $x=0.5$ and $\agt 5$ MeV for $x=0.3$.
Our result suggests the existence of ``spongelike'' phases
with negative Euler characteristic
in addition to the simple ``pasta'' phases
in supernova cores until $T \alt 3$ MeV.
\end{abstract}


\maketitle


\section{Introduction}

In the process of the collapse-driven supernova \cite{bethe},
matter in the core experiences adiabatic compression:
the central density increases from $\sim 10^{9}$ g cm$^{-3}$
at the beginning of the collapse
to around the normal nuclear density $\rho_{0}=0.165$ fm$^{-3}$
just before bounce;
the temperature reaches $\sim O(1)$ MeV at this point.

At subnuclear densities,
nuclear matter exhibits the coexistence of a liquid phase with a gas phase
due to the internucleon interaction which has an attractive part.
In the density region where nuclei are about to melt into uniform matter,
it is expected that, at sufficiently low temperatures
relevant to neutron star interiors \cite{review},
the energetically favorable configuration
of the mixed phase possesses interesting spatial structures
such as rodlike and slablike nuclei and rodlike and spherical bubbles, etc.,
which are referred to as nuclear ``pasta'' \cite{rpw,hashimoto}.

This prediction is confirmed by several approaches assuming nuclear shapes
such as the liquid drop models \cite{lorenz,gentaro1,gentaro2} and
the Thomas-Fermi calculations \cite{oyamatsu},
and is also confirmed without assuming nuclear shapes
in the framework of the Thomas-Fermi approximation \cite{williams}
and of the quantum molecular dynamics (QMD) \cite{qmd1,qmd2}.
While nuclear pasta at zero temperature is studied by several authors,
pasta phases at finite temperatures relevant to supernova inner cores
have not been studied yet
except for a work by Lassaut {\it et al.} using the Thomas-Fermi
approximation \cite{lassaut} and brief estimates of
thermal fluctuations of the long-wave-length mode \cite{gentaro1,gentaro2}.
It is noted that,
at temperatures of several MeV, effects of thermal fluctuations
on nucleon distribution would be quite significant at subnuclear densities.
However, the mean-field approximation such as the Thomas-Fermi \cite{tf}
and Hartree-Fock \cite{bonche} approximation is not suitable to
incorporate thermal fluctuations.

Finite temperature effects lead to
evaporation of nucleons from nuclear liquid region and
smoothed nucleon density profiles.
At lower temperatures where each nuclei fluctuates a little
around an average species, a compressible liquid-drop model \cite{liquid drop}
with incorporating the temperature dependence of
its bulk, surface, and Coulomb+lattice components
provides a useful way to investigate the pasta phases
at finite temperatures.
As for the bulk component \cite{bulk}, binding energy, saturation density,
and incompressibility \cite{incompress}, which are parameters characterizing
saturation properties, decrease with increasing temperature
while the temperature dependence of the symmetry energy is not significant
\cite{sym}.
For surface component, thermal broadening of the nucleon density profile
reduces nuclear surface tension \cite{surf}.
Lattice energy is also modified by translational motion of nuclei
\cite{liquid drop}.
However, at higher temperatures where the fluctuation of nuclear shape
is significant, the above liquid-drop picture no longer holds;
we have to call on some {\it ab-initio} method which does not assume nuclear shape.
More interestingly, it might be possible that, at these temperatures,
the shape of the nuclear surface fluctuates and
nuclei of various sizes and shapes coexist like colloid
due to the entropy effect.

These finite temperature effects can be well described
by the methods of molecular dynamics (MD) for nucleon many-body systems
(see, e.g., Ref.\ \cite{feldmeier} for review).
QMD \cite{aichelin}, which is one of them,
enables us to treat much larger systems
than the other methods of MD do.
Furthermore, at temperatures of several MeV,
shell effects, which cannot be incorporated by QMD,
are less important because they washed away by thermal fluctuations
above $\sim 3$ MeV.
Thus QMD is an efficient and trustable method for studying
nuclear matter at finite temperatures \cite{maruyama,peilert}.

Pasta phases in supernova matter (SNM) are expected to affect
the neutrino transport and hydrodynamics in supernova cores.
Let us first note that the neutrino wavelengths, typically of order 20 fm, 
are comparable to or even greater than the internuclear spacing,
leading to diffractive effects on the neutrino elastic scattering
off such a periodic spatial structure of nuclear matter \cite{rpw}.
These effects, induced by the internuclear Coulombic correlations,
would reduce the scattering rates and hence the lepton 
fraction $Y_{L}$.
For the bcc lattice of spherical nuclei, such a reduction was
examined by Horowitz \cite{horowitz} by calculating the 
associated static structure factor.
It is also noteworthy that 
nonspherical nuclei and bubbles are elongated in specific direction.
In such direction, the neutrino scattering processes are
no longer coherent,
in contrast to the case of
roughly spherical nuclei whose finiteness in any direction
yields constructive interference in the scattering,
which leads to the neutrino-trapping \cite{freedman,sato}.
The final point to be mentioned is that
the changes in the nuclear shape are accompanied by
discontinuities in the adiabatic index,
denoting how hard the
equation of state of the material is.
These discontinuities
may influence the core hydrodynamics during the initial phase of 
the collapse \cite{lassaut}.

In the present paper, we study the structure of hot dense matter
at subnuclear densities within the framework of QMD.
Simulations of nuclear matter with proton fraction $x=0.3$
in addition to symmetric nuclear matter ($x=0.5$) are performed
because the typical value of the proton fraction for supernova matter
is around 0.3 due to the trapping of the electron neutrinos
\cite{freedman,sato}.
We draw phase diagrams for $x=0.5$ and 0.3 in the density versus
temperature plane, which show the qualitative feature of the
finite temperature effects on the structure of nuclear matter.
The results of the present study would be helpful
to understand the real situation of the interior
of the collapsing cores.

The plan of this paper is as follows.
In Section \ref{formulation},
we briefly explain the QMD model used in the present study
and then discuss an effective temperature.
In this section, a thermostatting method used in the simulations
is also explained.
In Section \ref{results},
we show snapshots of some typical nucleon distributions
to discuss qualitative features of finite temperature effect.
After that, we analyze the structure of matter
using two-point correlation functions and Minkowski functionals,
and finally, resultant phase diagrams are shown.
Summary and conclusion are presented in Section \ref{conclusion}.

\section{Formulation\label{formulation}}

\subsection{Model Hamiltonian\label{model}}

Simulating nuclear matter at subnuclear densities within the framework of QMD,
we use a QMD model Hamiltonian developed by Maruyama {\it et al.} \cite{maruyama},
which is constructed so as to reproduce bulk properties of nuclear matter
and properties of finite nuclei.
This model Hamiltonian, which describes interactions between nucleons,
consists of the following six terms
\begin{equation}
  {\cal H} =
  T+V_{\rm Pauli}+V_{\rm Skyrme}+V_{\rm sym}+V_{\rm MD}+V_{\rm Coulomb}\ ,
  \label{hamiltonian}
\end{equation}
where $T$ is the kinetic energy,
$V_{\rm Pauli}$ is the Pauli potential introduced to reproduce
the Pauli principle effectively,
$V_{\rm Skyrme}$ is the Skyrme potential
which consists of an attractive two-body term and a repulsive three-body term,
$V_{\rm sym}$ is the symmetry potential,
$V_{\rm MD}$ is the momentum-dependent potential
introduced as two Fock terms of the Yukawa interaction and
$V_{\rm Coulomb}$ is the Coulomb energy including the constant contribution
$V_{p\mbox{-}e}$ due to the Coulomb interaction between protons and electrons
\cite{note vpe}.
The expressions of these terms are given as
\begin{widetext}
\begin{eqnarray}
  T  &=& \sum_{i, j(\ne i)} \frac{\bf P_{\it i}^{2}}{2 m_{i}}\ ,\label{kin}\\  
  V_{\rm Pauli} &=& 
  \frac{1}{2}\
  C_{\rm P}\left( \frac{\hbar}{q_0 p_0}\right)^3
  \sum_{i, j(\neq i)} 
  \exp{ \left [ -\frac{({\bf R}_i-{\bf R}_j)^2}{2q_0^2} 
          -\frac{({\bf P}_i-{\bf P}_j)^2}{2p_0^2} \right ] }\
  \delta_{\tau_i \tau_j} \delta_{\sigma_i \sigma_j}\ ,\label{pauli}\\
  V_{\rm Skyrme} &=&
  {\alpha\over 2\rho_0}\sum_{i, j (\neq i)}
  \rho_{ij}
  +  {\beta\over (1+\tau)\ \rho_0^{\tau}}
  \sum_i \left[ \sum_{j (\neq i)} \int { d^3{\bf r} \ \tilde{\rho_i}({\bf r}) \
                       \tilde{\rho_j}({\bf r}) } \right]^{\tau}\ ,
                   \label{skyrme}\\
  V_{\rm sym} &=&
  {C_{\rm s}\over 2\rho_0} \sum_{i , j(\neq i)} \,
  ( 1 - 2 | \tau_i - \tau_j | ) \ \rho_{ij}\ ,\label{sym}\\
  V_{\rm MD}  &=&
  V_{\rm MD}^{(1)} + V_{\rm MD}^{(2)} \nonumber \\
  &=&
         {C_{\rm ex}^{(1)} \over 2\rho_0} \sum_{i , j(\neq i)} 
      {1 \over 1+\left[{{\bf P}_i-{\bf P}_j \over \hbar \mu_1}\right]^2} 
      \ \rho_{ij}
     +   {C_{\rm ex}^{(2)} \over 2\rho_0} \sum_{i , j(\neq i)} 
      {1 \over 1+\left[{{\bf P}_i-{\bf P}_j \over \hbar \mu_2}\right]^2} 
      \ \rho_{ij}\ ,\label{md}\\
  V_{\rm Coulomb} &=&
  {e^2 \over 2}\sum_{i , j(\neq i)}
  \left(\tau_{i}+\frac{1}{2}\right) \, \left(\tau_{j}+\frac{1}{2}\right)
  \int\!\!\!\!\int d^3{\bf r}\,d^3{\bf r}^{\prime} 
  { 1 \over|{\bf r}-{\bf r}^{\prime}|} \,
  \rho_i({\bf r})\rho_j({\bf r}^{\prime})+V_{p\mbox{-}e}\ ,\label{coulomb}
\end{eqnarray}
\end{widetext}
where $\rho_{ij}$ means the overlap between the single-nucleon densities,
$\rho_{i}({\bf r})$ and $\rho_{j}({\bf r})$,
for $i$-th and $j$-th nucleons given as
\begin{equation}
  \rho_{ij} \equiv \int { d^3{\bf r} \ \rho_i({\bf r}) \
                       \rho_j({\bf r}) }\ ,
\end{equation}
$\sigma_{i}$ is the nucleon spin and $\tau_{i}$ is the isospin
($\tau_{i}=1/2$ for protons and $-1/2$ for neutrons) and
$C_{\rm P},\ q_{0},\ p_{0},\ \alpha,\ \beta,\ \tau,\ C_{\rm s},\ C_{\rm ex}^{(1)},\ C_{\rm ex}^{(2)},\ \mu_{1},\ \mu_{2}$
and $L$ are model parameters
determined to reproduce the properties of the ground states
of the finite nuclei, especially heavier ones,
and the saturation properties of nuclear matter \cite{maruyama}.
A parameter set used in this work is shown in Table \ref{parameter}.
The single-nucleon densities $\rho_{i}({\bf r})$
and $\tilde{\rho_{i}}({\bf r})$
are given as
\begin{eqnarray}
  \rho_i({\bf r}) & = & \left| \phi_{i}({\bf r}) \right|^{2}
  = \frac{1}{(2\pi L)^{3/2}}\ \exp{\left[
                - \frac{({\bf r} - {\bf R}_i)^2}{2L} \right]}\ ,\quad \label{packet}\\
  \tilde{\rho_i}({\bf r}) & = &
  \frac{1}{(2\pi \tilde{L})^{3/2}}\ \exp{\left[
                - \frac{({\bf r} - {\bf R}_i)^2}{2\tilde{L}} \right]}\ ,
\end{eqnarray}
with
\begin{equation}
  \tilde{L} = \frac{(1+\tau)^{1/ \tau}}{2}\ L\ .
\end{equation}
The quantity $\tilde{\rho_{i}}({\bf r})$
is introduced in the three body term of Skyrme interaction Eq. (\ref{skyrme})
to incorporate the effect of the repulsive density-dependent term
by the modified width $\tilde{L}$.

\begin{table}[h]
\caption{Effective interaction parameter set\\
  \qquad ($K$=280 MeV; medium EOS model in Ref.\ \cite{maruyama})}
\begin{ruledtabular}
\begin{tabular}{cccc}
& $\alpha$ (MeV) &\qquad\qquad $-92.86$ &\\
& $\beta$ (MeV) &\qquad\qquad 169.28 &\\
& $\tau$ &\qquad\qquad 1.33333 &\\
& $C_{\rm s}$ (MeV) &\qquad\qquad 25.0 &\\
& $C_{\rm ex}^{(1)}$ (MeV) &\qquad\qquad $-258.54$ &\\
& $C_{\rm ex}^{(2)}$ (MeV) &\qquad\qquad 375.6 &\\
& $\mu_1$ (MeV) &\qquad\qquad 2.35 &\\
& $\mu_2$ (MeV) &\qquad\qquad 0.4 &\\
& $L$ (fm$^2$) &\qquad\qquad 2.1 &\\
\end{tabular}
\end{ruledtabular}
\label{parameter}
\end{table}

\subsection{Effective temperature}

The effective Hamiltonian (\ref{hamiltonian}) used in this work
contains momentum dependent interactions,
i.e. the Pauli potential $V_{\rm Pauli}$ and
the momentum dependent potential $V_{\rm MD}$.
Thus the usual expression for the instantaneous kinetic temperature
$T_{\rm kin}$ given as
\begin{equation}
  \frac{3}{2} k_{\rm B} T_{\rm kin}
  = \frac{1}{N} \sum_{i=1}^{N} \frac{{\bf P}_{i}^{2}}{2m_{i}}\ ,\label{tkin}
\end{equation}
loses its meaning of the temperature in thermodynamics.
We here use an effective temperature $T_{\rm eff}$
proposed by Chikazumi {\it et al.} \cite{chikazumi},
which is given as
\begin{eqnarray}
  \frac{3}{2} k_{\rm B} T_{\rm eff}
  &=& \frac{1}{N} \sum_{i=1}^{N} \frac{1}{2}{\bf P}_{i} \cdot
  \frac{d {\bf R}_{i}}{dt}\nonumber\\
  &=& \frac{1}{N} \sum_{i=1}^{N} \frac{1}{2}{\bf P}_{i} \cdot
  \left(\frac{\partial {\cal H}}{\partial {\bf P}_{i}}\right)\ .\label{teff}
\end{eqnarray}
It can be immediately seen that
this expression is equivalent to the usual kinetic temperature
defined by Eq.\ (\ref{tkin})
if the effective Hamiltonian ${\cal H}$ does not have
momentum dependent interactions (i.e., $V_{\rm Pauli}$ and $V_{\rm MD}$).

\begin{figure}
\begin{center}
\rotatebox{0}{\hspace{-0.5cm}
\resizebox{9cm}{!}
{\includegraphics{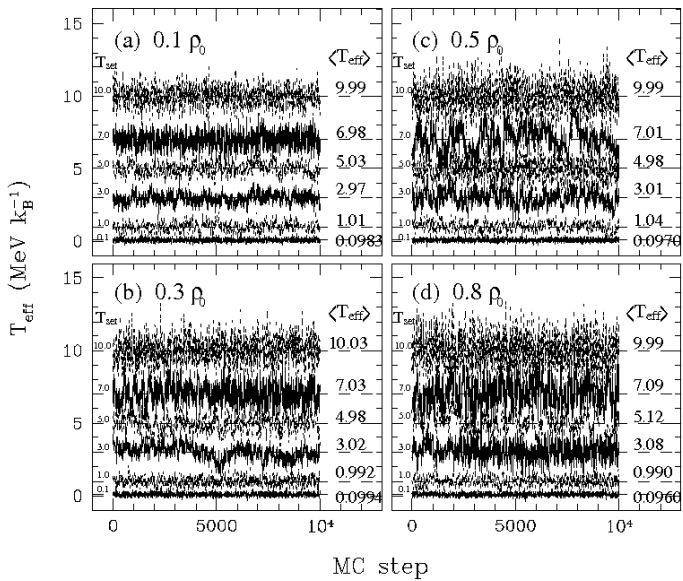}}}
\caption{\label{fig teff}
  Sampled values of the instantaneous effective temperature $T_{\rm eff}$
  and their long-time averages $\langle T_{\rm eff} \rangle$
  for $\rho=0.1,\ 0.3,\ 0.5$ and 0.8 $\rho_{0}$.
  }
\end{center}
\end{figure}

In order to confirm whether the effective temperature $T_{\rm eff}$ is
consistent with temperature in the Boltzmann statistics,
we perform Metropolis Monte Carlo (MC) simulations
\cite{metropolis,allen,frenkel}
with 256 nucleons (with $x=0.5$, i.e., 128 protons and 128 neutrons).
We investigate six given temperatures
$T_{\rm set}= 0.1,\ 1,\ 3,\ 5,\ 7$ and 10 MeV
at four nucleon densities $\rho = 0.1,\ 0.3,\ 0.5$ and 0.8 $\rho_{0}$
within a wide region of the phase diagram at subnuclear densities
covering from a phase-separating region to a uniform fluid region.
We prepare a cubic box,
which is imposed of the periodic boundary condition.
In the simulations
the system is equilibrated at a given temperature $T_{\rm set}$
for 1000 MC steps (i.e. $1000 \times N$ trial moves),
and then sampling is carried out for the following 10000 MC steps.
Sampled values of the instantaneous effective temperature $T_{\rm eff}$
is plotted in Fig.\ \ref{fig teff}.
We can see from this figure that the instantaneous effective temperature
$T_{\rm eff}$ fluctuates around the given value of $T_{\rm set}$.
It is noted that the long-time averaged values $\langle T_{\rm eff} \rangle$
of the effective temperature coincide with $T_{\rm set}$
quite well within the fluctuations of order $\sim T_{\rm set}/\sqrt{N}$
due to the finite particle number.
Thus we can conclude that
the effective temperature given by Eq.\ (\ref{teff}) is consistent
with temperature in the Boltzmann statistics.
It is also confirmed that,
in microcanonical molecular dynamics simulations,
the mean value of the effective temperature
keeps constant after the system is equilibrated enough.

The instantaneous effective temperature can be negative
as plotted in Fig.\ \ref{fig teff} for $T_{\rm set}=0.1$
when ${\bf P}_{i}$ and $\dot{{\bf R}}_{i}$ take the opposite directions
each other due to the contribution of the momentum dependent interactions.
However, it is also confirmed that,
after the system is relaxed,
the long-time average of the effective temperature
does not take negative values when we pursue the time evolution
of the system by the QMD equations of motion
even though some friction terms were attached to them like Eqs.\ (17)
in Ref.\ \cite{qmd2}.
In the remaining part of this paper, we measure temperature $T$
by the effective temperature.

\subsection{Thermostat \label{thermostat}}

It is necessary to perform ``isothermal'' QMD simulations
in order to equilibrate the system at a specified effective temperature.
In ordinary molecular dynamics simulations,
what is called the Nos\'e-Hoover thermostat is commonly used to carry out
constant-temperature simulations \cite{nose,hoover,allen,frenkel}.
The approach of Nos\'e and Hoover is based on the Hamiltonian
of an extended system,
which contains additional and artificial coordinates and velocities
intended to mimic the dynamics of the system in contact with a thermal bath.
It is shown by Nos\'e that
this method generates the states in the canonical ensemble average,
i.e., the microcanonical ensemble average in the extended system
reduces to the canonical ensemble average in the real system.
For example, the momentum distribution coincides with
the Maxwell-Boltzmann distribution exactly
after the system is well equilibrated with this thermostat.

Here we modify the Nos\'e-Hoover's method
so as to adapt to the effective temperature.
The Hamiltonian of the extended system in this case can be written as
\begin{eqnarray}
  {\cal H}_{\rm Nose} &=& \sum_{i=1}^{N} \frac{{\bf P}_{i}^{2}}{2m_{i}}
  + {\cal U}(\{{\bf R}_{i}\}, \{{\bf P}_{i}\})
  + \frac{s^2p_{s}^{2}}{2Q} + g \frac{\ln s}{\beta}\nonumber \\
  &=& {\cal H} + \frac{s^2p_{s}^{2}}{2Q} + g \frac{\ln s}{\beta}\ ,
\end{eqnarray}
where ${\cal U}(\{{\bf R}_{i}\}, \{{\bf P}_{i}\})$ is the potential
which depends on both positions and momenta,
$s$ is the additional dynamical variable for time scaling,
$p_{s}$ is the momentum conjugate to $s$,
$Q$ is the thermal inertial parameter
corresponding to a coupling constant between the system and thermostat
[in our simulations, we set $Q\sim 10^8$ MeV (fm/$c$)$^2$],
$g$ is a parameter to be determined as $3N$ by a condition
for generating the canonical ensemble
in the classical molecular dynamics simulations,
and $\beta$ is defined as $\beta \equiv 1/k_{\rm B}T_{\rm set}$.
The equations of motion yield
\begin{eqnarray}
  \frac{d {\bf R}_{i}}{dt}
  &=& \frac{\partial{\cal H}_{\rm Nose}}{\partial {\bf P}_{i}}
  = \frac{{\bf P}_{i}}{m_{i}}
  + \frac{\partial{\cal U}}{\partial {\bf P}_{i}}\ ,\\
  \nonumber\\
  \frac{d {\bf P}_{i}}{dt}
  &=& -\frac{\partial{\cal H}_{\rm Nose}}{\partial {\bf R}_{i}}
  = -\frac{\partial{\cal U}}{\partial{\bf R}_{i}} - \xi {\bf P}_{i}\ ,\\
  \nonumber\\
  \frac{1}{s} \frac{ds}{dt}
  &=& \frac{1}{s} \frac{\partial{\cal H}_{\rm Nose}}{\partial p_{s}}
  = \frac{1}{Q} \frac{\partial{\cal H}_{\rm Nose}}{\partial\xi} = \xi\ ,\\
  \nonumber\\
  \frac{d\xi}{dt}
  &=& \frac{1}{Q} \left\{ \sum_{i=1}^{N}\left(
      \frac{{\bf P}_{i}^{2}}{m_{i}}
      + {\bf P}_{i}\cdot\frac{\partial{\cal U}}{\partial{\bf P}_{i}}
      \right) - \frac{g}{\beta} \right\},\label{dxi}
\end{eqnarray}
with
\begin{equation}
  \xi \equiv \frac{sp_{s}}{Q}\ ,
\end{equation}
where $\xi$ means the thermodynamic friction coefficient.

In the dynamical process described by these equations,
${\cal H}_{\rm Nose}$ is conserved and
the value of the effective temperature fluctuates
around $T_{\rm set}$ as can be seen from Eq.\ (\ref{dxi}).

\section{Simulations and Results\label{results}}

\subsection{Procedures for simulations\label{procedure}}

Let us here explain the procedure for the simulations.
We investigate nuclear matter with proton fractions $x=0.5$ and 0.3
at subnuclear densities in sufficiently wide regions
of the density versus temperature plane covering the whole
region where phase separation is observed:
symmetric nuclear matter is studied up to
$\rho=0.7\rho_{0}$ and $T=8$ MeV, and
nuclear matter with $x=0.3$, up to $\rho=0.6\rho_{0}$ and $T=7$ MeV.
Intervals of the density and the temperature between the investigated points 
are $0.025\rho_{0}$ or $0.05\rho_{0}$ and 0.5 MeV or 1 MeV, respectively
(from the present section onward, we set $k_{\rm B}=1$).

We perform simulations for a cubic box with periodic boundary condition.
We study the $(n,p,e)$ system with 2048 nucleons
(for some typical cases of the columnar phase and the planar phase,
a system with 16384 nucleons is also used).
Throughout this paper, we treat systems which are not magnetically polarized,
i.e., they contain equal numbers of protons (and neutrons) with spin up and
spin down.
The relativistic degenerate electrons which ensure charge neutrality
are regarded as a uniform background because the influence of 
the electron screening on the phase diagram at subnuclear densities is small
as shown explicitly in Ref.\ \cite{screening}.
The Coulomb interaction is calculated by the Ewald method
taking account of the Gaussian charge distribution of the proton wave packets
(see, e.g., Appendix A in Ref.\ \cite{qmd2}),
which enables us to sum up the contributions of long-range interactions
in a system with periodic boundary conditions efficiently.
For the nuclear interaction, we use the effective Hamiltonian
developed by Maruyama {\it et al.} (medium EOS model) \cite{maruyama},
whose expressions are given in Section \ref{model}.

We first prepare a hot, uniform gas with 2048 nucleons
at $T \sim 20$ MeV as an initial condition,
which is equilibrated for $\sim 500 - 2000$ fm/$c$ in advance.
We then cool it down slowly for $O(10^{3}-10^{4})$ fm/$c$
keeping the nucleon density unchanged
by the frictional relaxation method [see Eqs.\ (17) of Ref.\ \cite{qmd2}]
until the temperature reaches $\sim 5$ MeV. For the present QMD model,
this is the typical temperature
for the boundary of the phase-separating region at subnuclear densities
relevant to the pasta phases.
In some cases, the thermostat of the Nos\'e-Hoover type 
(see Section \ref{thermostat})
is also used to cool the system down quickly until $\simeq 10$ MeV,
at which temperature matter is still completely uniform.

After the cooling process,
the system is then relaxed for $\sim 4000-5000$ fm/$c$
at a given temperature $T_{\rm set}$
using the thermostat of the Nos\'e-Hoover type,
which is followed by a further relaxation for $\sim 5000$ fm/$c$
at the same $T_{\rm set}$ without the thermostat 
(i.e., microcanonical molecular dynamics simulation).
Thermal averages are measured in the microcanonical relaxation process.
The above relaxation processes with and without the thermostat
are repeated for the other values of $T_{\rm set}$
by changing $T_{\rm set}$ by 0.5 or 1 MeV and keeping the density constant.

Simulations of a larger system with 16384 nucleons have also been performed
for some typical cases of the phases with slablike nuclei 
and with rodlike nuclei to examine the importance of finite size effects.
We combine eight replicated samples at $T=0$ with 2048 nucleons
into a 16384-nucleon sample.
We then add numerical noise to the positions and the momenta of nucleons,
up to 0.1 fm in the position and 1 MeV/$c$ in the momentum.
We increase the temperature by 1MeV
and relax the system for $\sim 4000-5000$ fm/$c$
using the Nos\'e-Hoover thermostat and relax further 
for $\sim 3000-5000$ fm/$c$ without the thermostat.
These relaxation processes are repeated for $T_{\rm set}=2$ and 3 MeV.

The simulations of the 2048-nucleon system are performed
using PCs (Pentium III) equipped with MDGRAPE-2,
and those of the 16384-nucleon system are done by
Fujitsu VPP 5000 equipped with MDGRAPE-2.

\subsection{Two-point correlation functions and 
Minkowski functionals\label{2pcf minko}}

To analyze the spatial distribution of nucleons,
we use the two-point correlation function.
The two-point correlation function $\xi_{ii}$
for the nucleon density field $\rho^{(i)}$ $(i=N,p,n;$ where $N$ stands for
nucleons$)$ is here defined as
\begin{eqnarray}
  \xi_{ii}(r)&=&
    \frac{1}{4\pi} \int d\Omega_{\bf r}\ \frac{1}{V} \int d^{3}{\bf x}\
    \delta_{i}({\bf x}) \delta_{i}({\bf x+r})\\
    &\equiv& \langle \delta_{i}({\bf x}) \delta_{i}({\bf x+r})
    \rangle_{{\bf x}, \Omega_{\bf r}}\ ,
\end{eqnarray}
where $\langle\cdots\rangle_{{\bf x}, \Omega_{\bf r}}$ denotes
an average over the position ${\bf x}$ and the direction of ${\bf r}$,
and $\delta_{i}({\bf x})$ is the fluctuation of the density field
$\rho^{(i)}({\bf x})$ given by
\begin{equation}
  \delta_{i}({\bf x}) \equiv \frac{\rho^{(i)}({\bf x}) - \overline{{\rho}^{(i)}}}
  {\overline{\rho^{(i)}}}\ ,
\end{equation}
with
\begin{equation}
  {\overline{\rho^{(i)}}} \equiv \frac{N_{i}}{V}\ .
\end{equation}

To identify the nuclear surface and
extract its morphological characteristics,
we use the Minkowski functionals
(see, e.g., Ref. \cite{minkowski} and references therein;
a concise review is provided by Ref.\ \cite{jens};
a brief explanation is given in Section IV C of Ref.\ \cite{qmd2}),
especially of the integral mean curvature
and the Euler characteristic \cite{genus}.

Suppose we set a threshold density $\rho_{\rm th}$ and
consider the regions where the density is higher than this value
surrounded by the isodensity surfaces for $\rho_{\rm th}$
(the procedure for identifying the nuclear surface,
which is characterized by isodensity surfaces for
a specific value of $\rho_{\rm th}$,
will be explained in Section \ref{phase diagrams}).
The integral mean curvature and the Euler characteristic
are defined as surface integrals of the following local quantities:
the mean curvature $H = (\kappa_{1}+\kappa_{2})/2$ and
the Gaussian curvature $G = \kappa_{1} \kappa_{2}$,
i.e., $\int_{\partial K} H dA$ and
$\chi \equiv \frac{1}{2 \pi} \int_{\partial K} G dA$,
where $\kappa_{1}$ and $\kappa_{2}$ are the principal curvatures and
$dA$ is the area element of the surface of the body $K$.
The Euler characteristic $\chi$ is a purely topological quantity
and is expressed as
\begin{eqnarray}
  \chi & = & \mbox{(number of isolated regions)}
  - \mbox{(number of tunnels)} \nonumber\\
  && + \mbox{(number of cavities)}.
  \label{euler}
\end{eqnarray}
Here we introduce their normalized quantities:
the area-averaged mean curvature,
$\langle H \rangle \equiv \frac{1}{A}\int H dA$,
and the Euler characteristic density, $\chi / V$,
where $V$ is the volume of the whole space.

In the present work, we use $64^3$ ($128^3$)
grid points for the 2048-nucleon (16384-nucleon) system
in constructing the nucleon density distribution $\rho^{(i)}({\bf x})$.
Detailed procedures for calculating $\xi_{ii}$ 
($\langle H \rangle$ and $\chi/V$) are given in
Section IV A (IV C) of Ref.\ \cite{qmd2}.

\subsection{Typical nucleon distributions for the phases with
rodlike and slablike nuclei\label{snapshots}}

\begin{figure*}[htbp]
\begin{center}
\resizebox{17cm}{!}
{\includegraphics{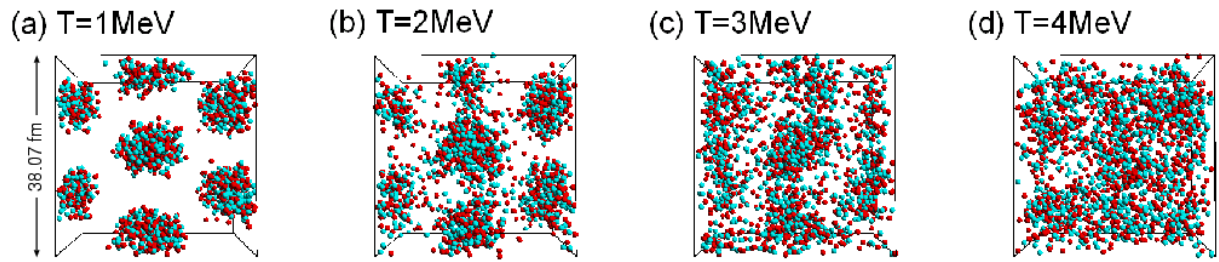}}
\caption{\label{snap 0.225rho x0.5 2000} (Color online)\quad
  The nucleon distributions for $x=0.5$, $\rho=0.225\rho_{0}$ at
  the temperatures of 1,2,3 and 4 MeV. 2048 nucleons are contained
  in the simulation box of size $L_{\rm box}=38.07$ fm.
  These figures show the top views along the axis of the rodlike nuclei.
  Protons are represented by the red particles, and
  neutrons by the green ones.
  }
\end{center}
\end{figure*}
\begin{figure*}[htbp]
\begin{center}
\resizebox{17cm}{!}
{\includegraphics{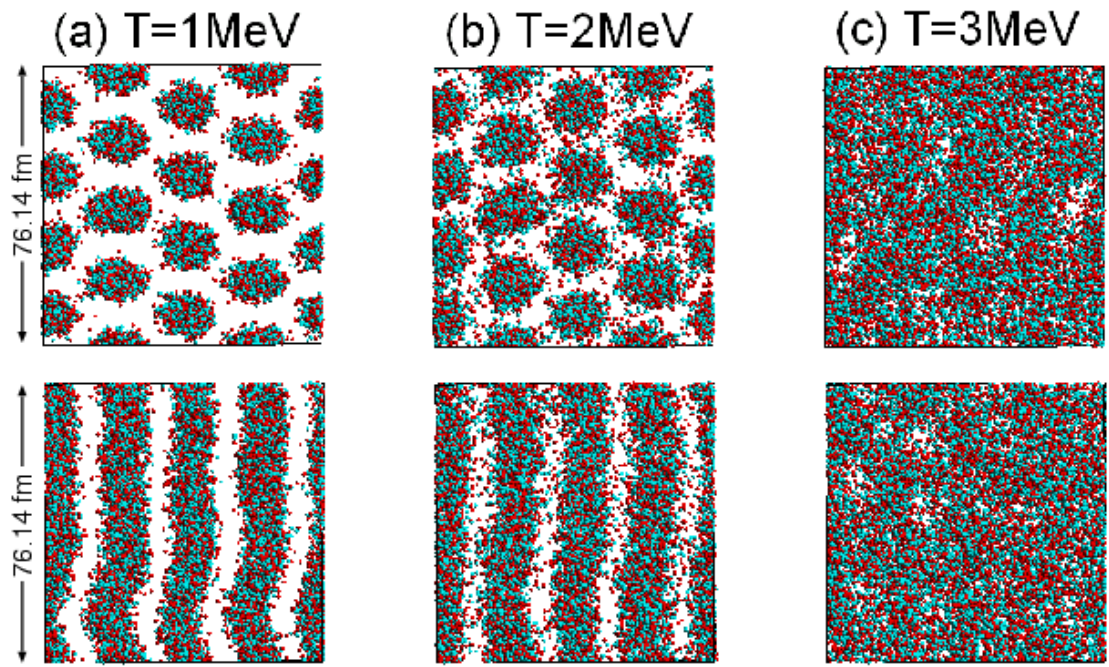}}
\caption{\label{snap 0.225rho x0.5 16000} (Color online)\quad
  The same as Fig.\ \ref{snap 0.225rho x0.5 2000} 
  at the temperatures of 1,2 and 3MeV for the system with
  16384 nucleons. The box size $L_{\rm box}$ is 76.14 fm.
  The upper panels show the top views along the axis of 
  the cylindrical nuclei at $T=0$,
  the lower ones the side views.
  }
\end{center}
\end{figure*}

\begin{figure*}[htbp]
\begin{center}
\resizebox{17cm}{!}
{\includegraphics{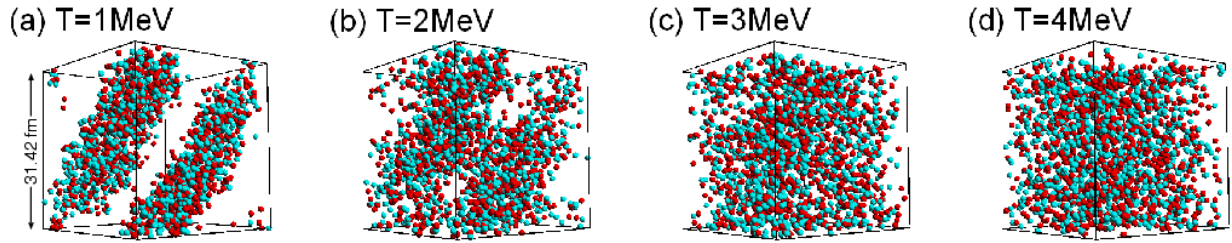}}
\caption{\label{snap 0.4rho x0.5 2000} (Color online)\quad
  The nucleon distributions for $x=0.5$, $\rho=0.4\rho_{0}$ at
  the temperatures of 1,2,3 and 4 MeV. 2048 nucleons are contained
  in the simulation box of size $L_{\rm box}=31.42$ fm.
  Protons are represented by the red particles, and
  neutrons by the green ones.
  }
\end{center}
\end{figure*}
\begin{figure*}[htbp]
\begin{center}
\resizebox{17cm}{!}
{\includegraphics{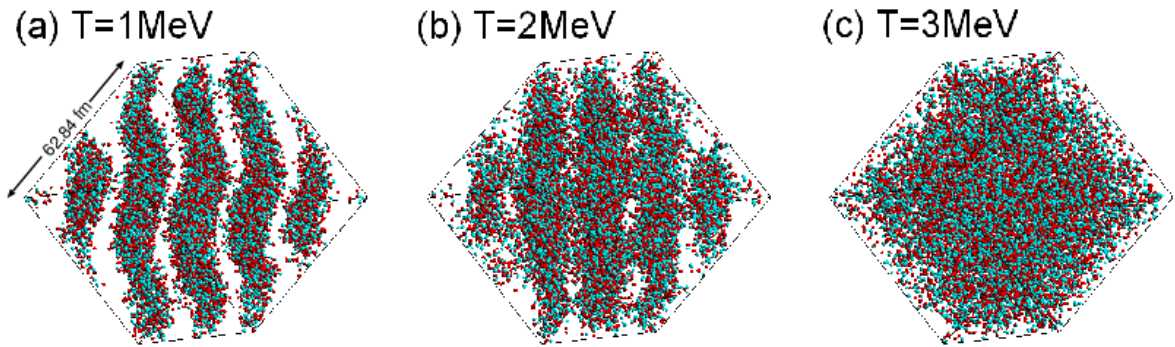}}
\caption{\label{snap 0.4rho x0.5 16000} (Color online)\quad
  The same as Fig.\ \ref{snap 0.4rho x0.5 2000} 
  at the temperatures of 1,2 and 3MeV for the system with
  16384 nucleons. The box size $L_{\rm box}$ is 62.84 fm.
  These figures are shown in the direction parallel
  to the plane of the slablike nuclei at $T=0$.
  }
\end{center}
\end{figure*}

Let us first show some snapshots of the nucleon distribution 
at finite temperatures for densities corresponding to
the phases with slablike nuclei and with rodlike nuclei at $T=0$.
These snapshots help us to understand
the qualitative feature of finite temperature effects on
the nuclear structure.

\begin{figure*}[htbp]
\begin{center}
\resizebox{17cm}{!}
{\includegraphics{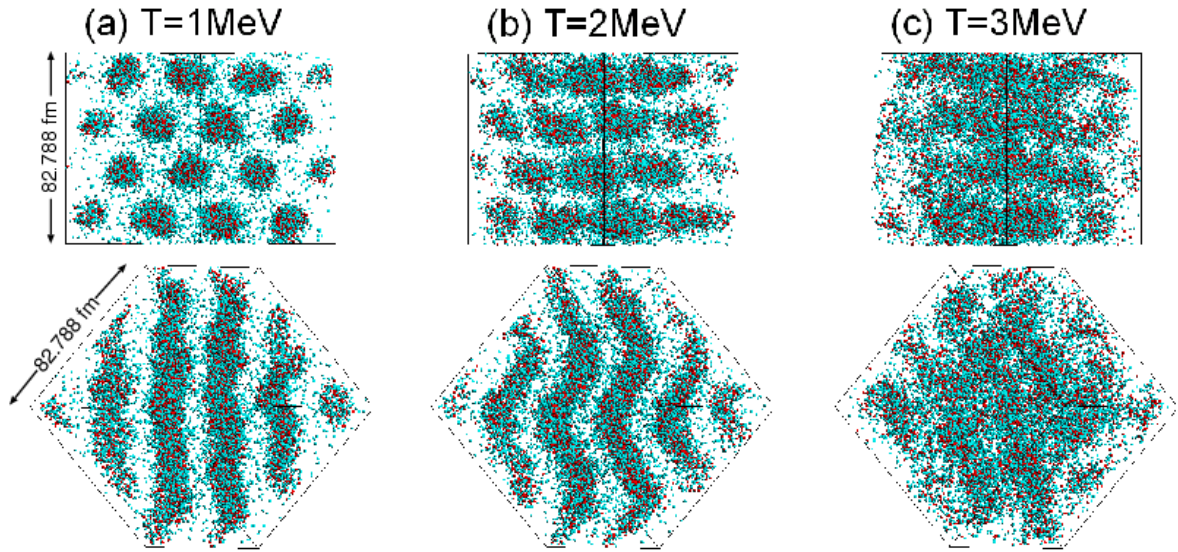}}
\caption{\label{snap 0.175rho x0.3 16000} (Color online)\quad
  The nucleon distributions for $x=0.3$, $\rho=0.175\rho_{0}$ at
  the temperatures of 1,2, and 3 MeV. 16384 nucleons are contained
  in the simulation box of size $L_{\rm box}=82.788$ fm.
  Protons are represented by the red particles, and
  neutrons by the green ones.
  The upper panels show the top views along the axis of 
  the cylindrical nuclei at $T=0$,
  the lower ones the side views.
  }
\end{center}
\end{figure*}

\begin{figure*}[htbp]
\begin{center}
\resizebox{17cm}{!}
{\includegraphics{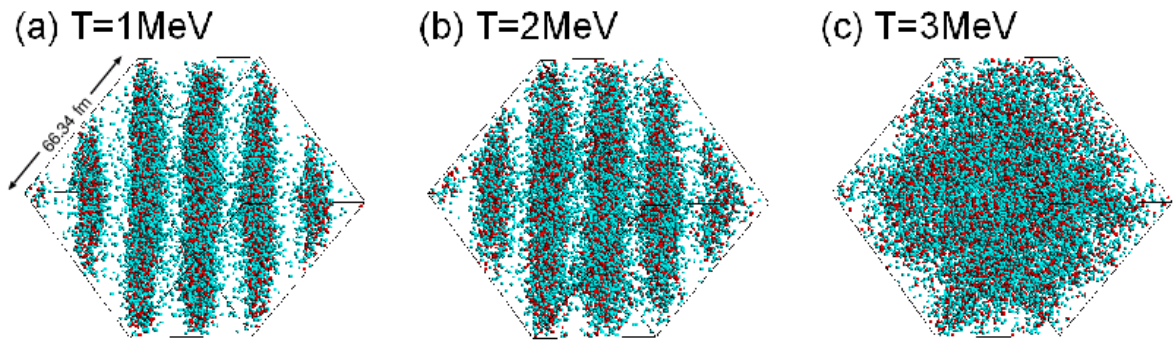}}
\caption{\label{snap 0.34rho x0.3 16000} (Color online)\quad
  The nucleon distributions for $x=0.3$, $\rho=0.34\rho_{0}$ at
  the temperatures of 1,2, and 3 MeV. 16384 nucleons are contained
  in the simulation box of size $L_{\rm box}=66.34$ fm.
  Protons are represented by the red particles, and
  neutrons by the green ones.
  These figures are shown in the direction parallel
  to the plane of the slablike nuclei at $T=0$.
  }
\end{center}
\end{figure*}

Figures \ref{snap 0.225rho x0.5 2000} and \ref{snap 0.225rho x0.5 16000} 
show snapshots of the
nucleon distribution for $x=0.5$ at a density of $0.225\rho_{0}$
(the phase with cylindrical nuclei at $T=0$)
and Figs.\ \ref{snap 0.4rho x0.5 2000} and \ref{snap 0.4rho x0.5 16000}
show those for $x=0.5$ and $\rho=0.4\rho_{0}$ 
(the phase with planar nuclei at $T=0$).
Here we show snapshots of the 2048-nucleon system and those of the
16384-nucleon system for both cases.
Figures \ref{snap 0.175rho x0.3 16000} and \ref{snap 0.34rho x0.3 16000}
for $x=0.3$ are the same as Figs.\ \ref{snap 0.225rho x0.5 16000}
and \ref{snap 0.4rho x0.5 16000} for $x=0.5$, respectively;
Fig.\ \ref{snap 0.175rho x0.3 16000} is for $0.175\rho_{0}$ 
(the phase with cylindrical nuclei at $T=0$), and
Fig.\ \ref{snap 0.34rho x0.3 16000} is for $0.34\rho_{0}$ 
(the phase with planar nuclei at $T=0$).
The snapshots for the 16384-nucleon systems 
(Figs.\ \ref{snap 0.225rho x0.5 16000}, \ref{snap 0.4rho x0.5 16000},
\ref{snap 0.175rho x0.3 16000}, and \ref{snap 0.34rho x0.3 16000})
are depicted without perspective.

\begin{figure}[htbp]
\begin{center}
\resizebox{8cm}{!}
{\includegraphics{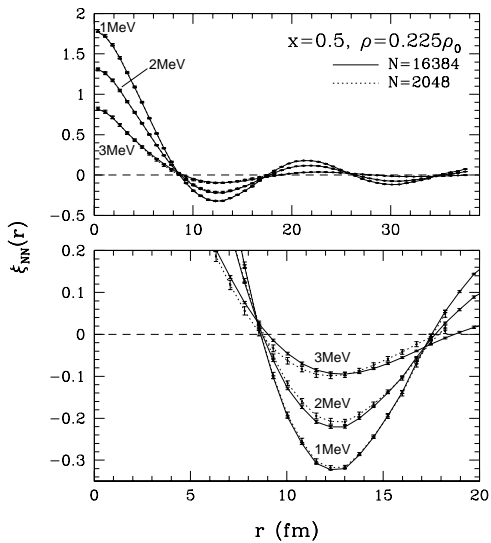}}
\caption{\label{corr2 0.225rho x0.5 compare}
  Comparison of the two-point correlation functions $\xi_{NN}$ 
  for the systems with 2048 and 16384 nucleons.
  These are calculated for $x=0.5$ and $\rho=0.225\rho_0$,
  where the system is in the phase with rodlike nuclei at zero temperature.
  The error bars are the standard deviations in the long-time average.
  }
\end{center}
\end{figure}

\begin{figure}[htbp]
\begin{center}
\resizebox{8cm}{!}
{\includegraphics{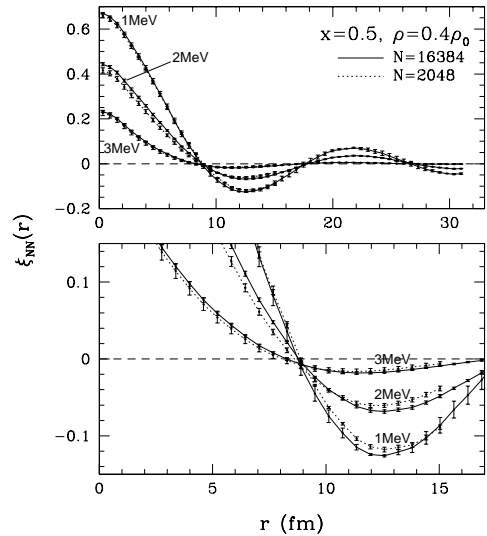}}
\caption{\label{corr2 0.4rho x0.5 compare}
  The same as Fig.\ \ref{corr2 0.225rho x0.5 compare}
  for $x=0.5$ and $\rho=0.4\rho_0$,
  where the system is in the phase with rodlike nuclei at zero temperature.
  }
\end{center}
\end{figure}

From these figures, we can see the following qualitative features 
irrespective of the proton fraction and the system size:
at $T\simeq 1.5-2$ MeV
(but snapshots for $T\simeq 1.5$ MeV are not shown there),
the number of the evaporated nucleons 
starts to be significant; at $T\agt 3$ MeV, 
the density profiles of the nucleons are smoothed out and 
it is difficult to identify the nuclear surface.
In view of the fact that these general features are the same
for systems with different particle number 
(see Figs.\ \ref{snap 0.225rho x0.5 2000} and \ref{snap 0.225rho x0.5 16000}
for $\rho=0.225\rho_0$; Figs.\ \ref{snap 0.4rho x0.5 2000} and
\ref{snap 0.4rho x0.5 16000} for $\rho=0.4\rho_0$),
we can say that a qualitatively correct phase diagram can be obtained
by using 2048-nucleon system.
This statement is supported by the behaviors of 
the two-point correlation function for these two systems.
It is remarkable that $\xi_{NN}$ for the 2048-nucleon system
and that for 16384-nucleon one coincide quite well,
as shown in Figs.\ \ref{corr2 0.225rho x0.5 compare} 
and \ref{corr2 0.4rho x0.5 compare}.
However, we should note that the larger system can incorporate
thermal fluctuations of longer wavelengths \cite{note fluctuation}.
As can be seen by comparing Figs.\ \ref{snap 0.4rho x0.5 2000}(a) and
\ref{snap 0.4rho x0.5 16000}(a), the slablike nuclei
have waves in the 16384-nucleon system at $T=1$ MeV,
but they do not in that with 2048 nucleons.

By comparing the two cases of different values of the proton fraction
in more detail, we can see that
the number of the evaporated protons is significantly smaller
than that of neutrons at $T \simeq 2$MeV for $x=0.3$
although they are close to each other at $T \simeq 2$MeV for $x=0.5$.
Here, the number density of the evaporated neutrons is defined as 
the number density of neutrons outside nuclei minus that of dripped neutrons
at $T=0$.
A nuclear matter region with higher proton fraction 
(but less than 0.5) is 
more energetically favorable than that with lower proton fraction
because of the symmetry energy, and
thus, at $x=0.3$, neutrons are preferentially evaporated
to increase the proton fraction in the nuclei.

We also note that, for $x=0.5$,
the slablike nuclei touch and fuse with each other
at $T=2$ MeV [see Figs.\ \ref{snap 0.4rho x0.5 2000}(b)
and \ref{snap 0.4rho x0.5 16000}(b)] 
while the rodlike structure persists at this temperature
[see Figs.\ \ref{snap 0.225rho x0.5 2000}(b)
and \ref{snap 0.225rho x0.5 16000}(b)].
The fragility of the phase with slablike nuclei would stem from
the Landau-Peierls instability and
its larger volume fraction for the nuclear matter region.

In closing the present section, we would like to mention
the effect of the dripped neutrons on the nuclear structure.
Here we note that
the dripped neutrons, on the one hand, suppress the thermal expansion of nuclei
due to its pressure acting on the nuclear surface,
and they, on the other hand, reduce the nuclear surface tension.
As can be seen by comparing
Figs.\ \ref{snap 0.225rho x0.5 16000}(a), (b) and
\ref{snap 0.175rho x0.3 16000}(a), (b),
rodlike nuclei tend to buckle at $x=0.3$ as the temperature increases,
but they do not at $x=0.5$ while just expand in radius.
At a fixed density and a fixed number of nuclei,
the nuclear radius is directly related to the volume fraction
of the nuclear matter region, which is principally determined by
the bulk properties.
Conversely, bending of nuclei is controled by elastic constants of nuclei,
which depend on nuclear surface tension and the Coulomb energy \cite{pp}.
The reduction of nuclear surface tension due to the dripped neutrons
leads to decrease of the elastic constants, which makes rodlike nuclei
easy to buckle in the case of $x=0.3$.
In the case of $x=0.5$, there is no reduction of the elastic constants
due to the dripped neutrons, and we thus only observe the thermal expansion
in radius of the rodlike nuclei.
Next, let us consider about slablike nuclei.
By comparing Figs.\ \ref{snap 0.4rho x0.5 16000}(b) and
\ref{snap 0.34rho x0.3 16000}(b), we can see that, for $x=0.5$,
the slablike nuclei expand in thickness and they touch with each other
at $T \sim 2$ MeV; for $x=0.3$, their expansion is smaller and
they do not touch at this temperature.
This result can be understood as a consequence of
suppressing the thermal expansion of nuclei due to the dripped neutrons.

\subsection{Phase diagrams\label{phase diagrams}}

In constructing phase diagrams, 
we determine the phase separating region,
identify the nuclear surface and classify the nuclear structure
according to its morphological characteristics.
For these purpose, we use the two-point correlation functions
and the Minkowski functionals introduced in Section \ref{2pcf minko}.

\begin{figure}[htbp]
\begin{center}
\resizebox{8cm}{!}
{\includegraphics{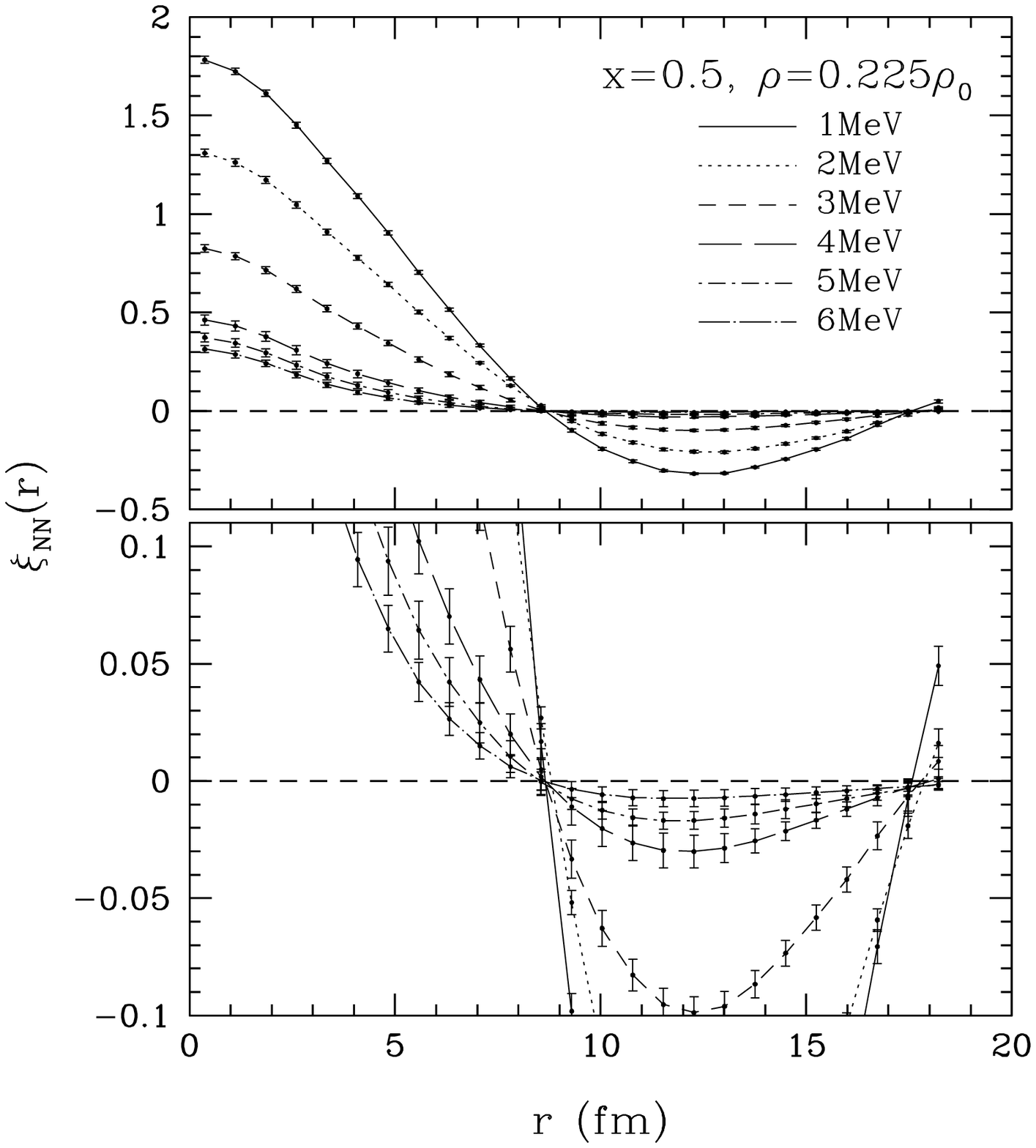}}
\caption{\label{corr2 0.225rho x0.5}
  Two-point correlation function of the density fluctuation
  calculated for $N=2048$, $x=0.5$ and $\rho=0.225\rho_0$,
  where the system is in the phase with rodlike nuclei at zero temperature.
  The error bars are the standard deviations in the long-time average.
  }
\end{center}
\end{figure}

\begin{figure}[htbp]
\begin{center}
\resizebox{8cm}{!}
{\includegraphics{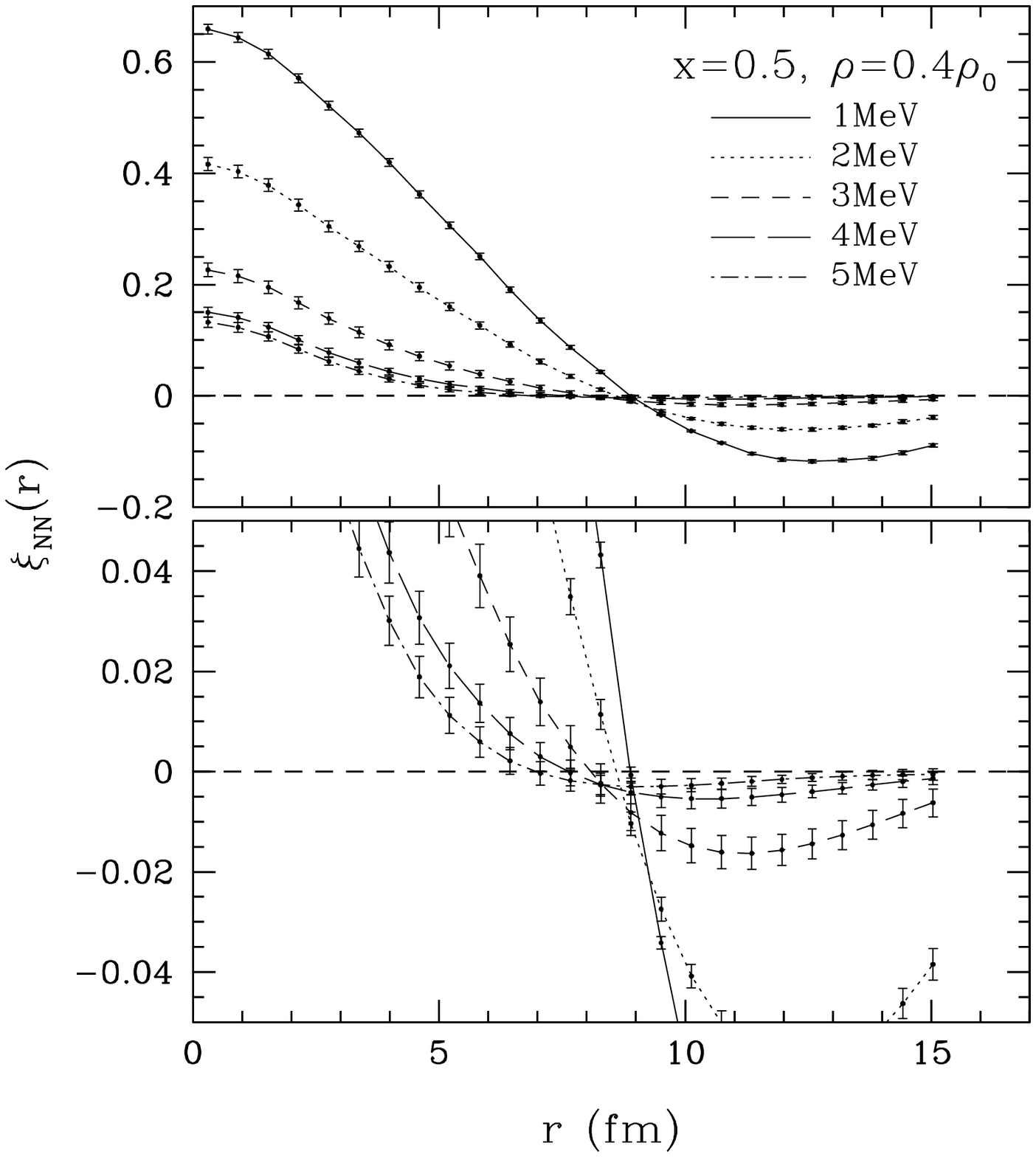}}
\caption{\label{corr2 0.4rho x0.5}
  Two-point correlation function of the density fluctuation
  calculated for $N=2048$, $x=0.5$ and $\rho=0.4\rho_0$,
  where the system is in the phase with slablike nuclei at zero temperature.
  The error bars are the standard deviations in the long-time average.
  }
\end{center}
\end{figure}

\begin{figure}[htbp]
\begin{center}
\resizebox{8cm}{!}
{\includegraphics{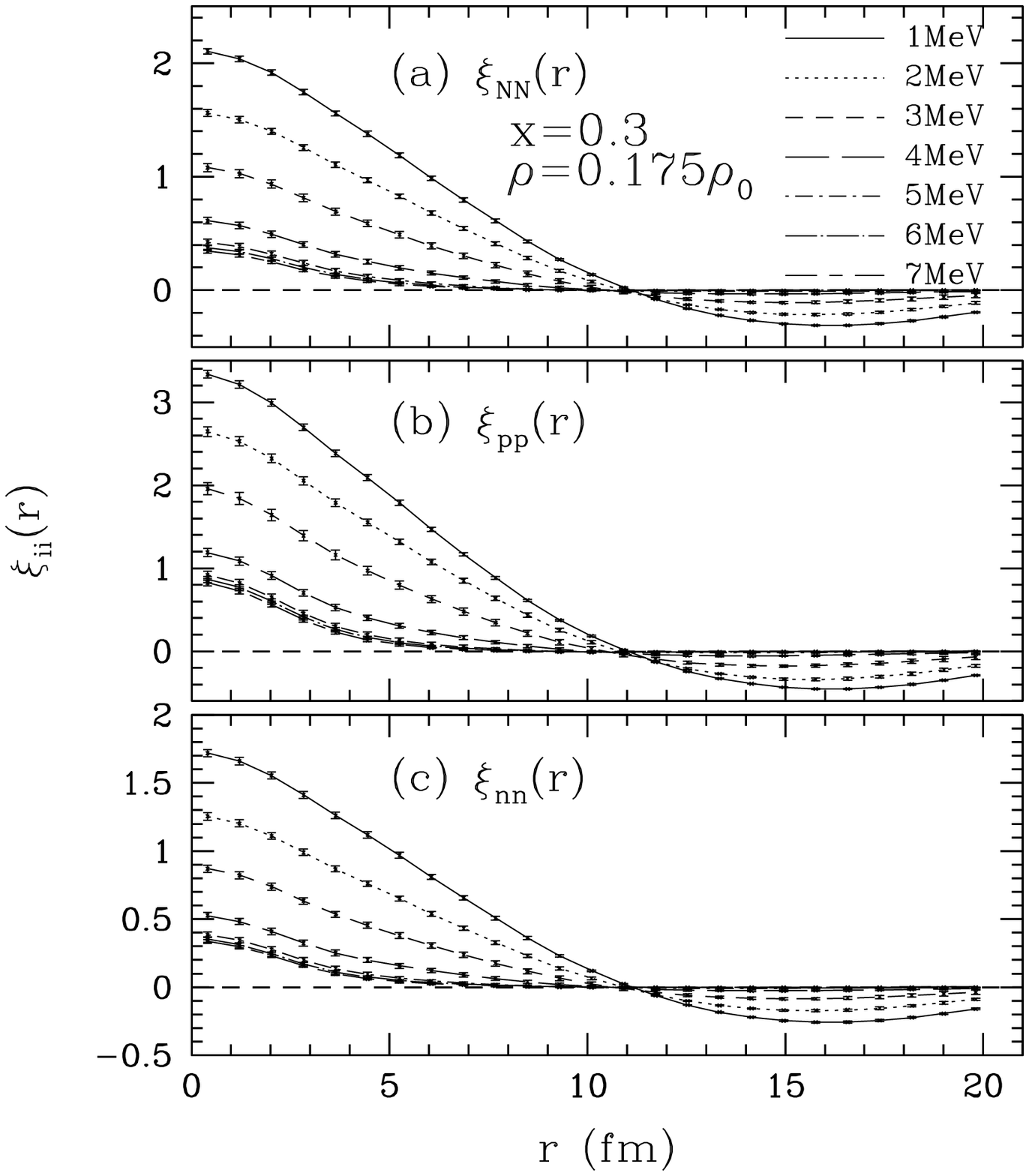}}
\caption{\label{corr2 0.175rho x0.3}
  Two-point correlation function of the density fluctuation
  calculated for $N=2048$, $x=0.3$ and $\rho=0.175\rho_0$,
  where the system is in the phase with rodlike nuclei at zero temperature.
  The error bars are the standard deviations in the long-time average.
  }
\end{center}
\end{figure}

\begin{figure}[htbp]
\begin{center}
\resizebox{8cm}{!}
{\includegraphics{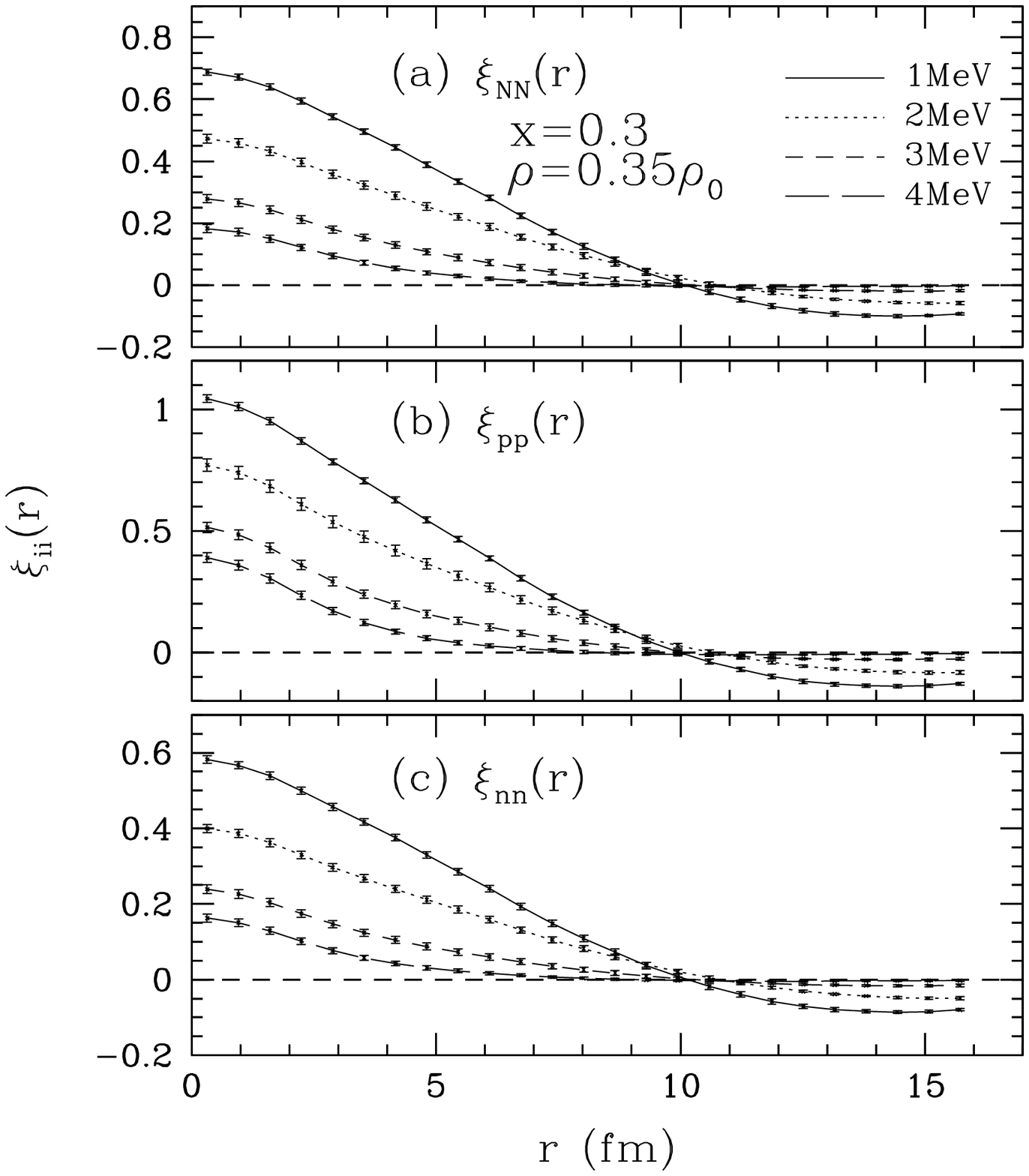}}
\caption{\label{corr2 0.35rho x0.3}
  Two-point correlation function of the density fluctuation
  calculated for $N=2048$, $x=0.3$ and $\rho=0.35\rho_0$,
  where the system is in the phase with slablike nuclei at zero temperature.
  The error bars are the standard deviations in the long-time average.
  }
\end{center}
\end{figure}

The phase-separation region is determined 
by the two-point correlation function $\xi_{NN}$
of the nucleon density fluctuation.
In Figs.\ \ref{corr2 0.225rho x0.5}, and \ref{corr2 0.4rho x0.5},
we plot $\xi_{NN}(r)$ for symmetric nuclear matter 
at $\rho=0.225$ and $0.4\rho_{0}$,respectively,
as typical examples.
The former density corresponds to the phase with rodlike nuclei
and the latter one to that with slablike nuclei at zero temperature
(see Figs.\ \ref{snap 0.225rho x0.5 2000} and \ref{snap 0.4rho x0.5 2000}).
From these figures, we can see that the amplitude of $\xi_{NN}$
and the relative density dispersion given by $\xi_{NN}(0)$ decrease
as the temperature increases.
It is also noted that the smallest value of $r$ at which $\xi_{NN}$ is zero
does not change much compared to the change in the amplitude.
This behavior of $\xi_{NN}$ is similar to that
when the density increases keeping $T=0$,
which we have studied in detail in Ref.\ \cite{qmd2}.
The important point is that, from the behavior of $\xi_{NN}$,
we can determine the temperature at which the long-range correlation
of the nucleon distributions disappears.
Figure \ref{corr2 0.225rho x0.5} (\ref{corr2 0.4rho x0.5})
shows that the long-range correlation
cannot be seen at $T=6$ MeV (5 MeV) and thus the phase-separating boundary
lies between $T=5$ and 6 MeV at $0.225\rho_{0}$
($T=4$ and 5 MeV at $0.4\rho_{0}$).

The two-point correlation functions $\xi_{ii}$ 
($i=N,p$, and $n$) for $x=0.3$ are also plotted
in Figs.\ \ref{corr2 0.175rho x0.3} and \ref{corr2 0.35rho x0.3}
for $\rho=0.175$ and $0.35\rho_{0}$, respectively.
As shown in Figs.\ \ref{snap 0.175rho x0.3 16000}
and \ref{snap 0.34rho x0.3 16000}, these densities correspond to
the phases with rodlike nuclei and with slablike nuclei
(Although the snapshots shown in these figures are 
for the 16384-nucleon system, the correlations functions $\xi_{ii}$
shown in Figs.\ \ref{corr2 0.175rho x0.3} and \ref{corr2 0.35rho x0.3}
are calculated for the 2048-nucleon system.).
The amplitude of $\xi_{ii}$ and the relative density dispersion
decrease with increasing temperature as in the case of $x=0.5$.
Among $\xi_{NN},\ \xi_{pp}$, and $\xi_{nn}$, 
the second one has the largest amplitude and the third one has the smallest
except for high temperatures at which matter is almost uniform.
We note that the smallest zero-points of the three kinds of $\xi_{ii}$
almost coincide each other and they remain nearly constant
at lower temperatures.
This behavior shows that the density fluctuations of protons
and neutrons are strongly correlated even at $x=0.3$
and the wavelength of the density fluctuations does not change significantly
at these temperatures.
As can be seen from Fig.\ \ref{corr2 0.175rho x0.3}(a)
[\ref{corr2 0.35rho x0.3}(a)], the long-range correlation disappears
at $T\ge5$ MeV [$T=4$ MeV] and thus the phase-separating boundary
lies between $T=4$ and 5 MeV at $0.175\rho_{0}$
[$T=3$ and 4 MeV at $0.35\rho_{0}$].

\begin{figure}[htbp]
\begin{center}
\resizebox{8.5cm}{!}
{\includegraphics{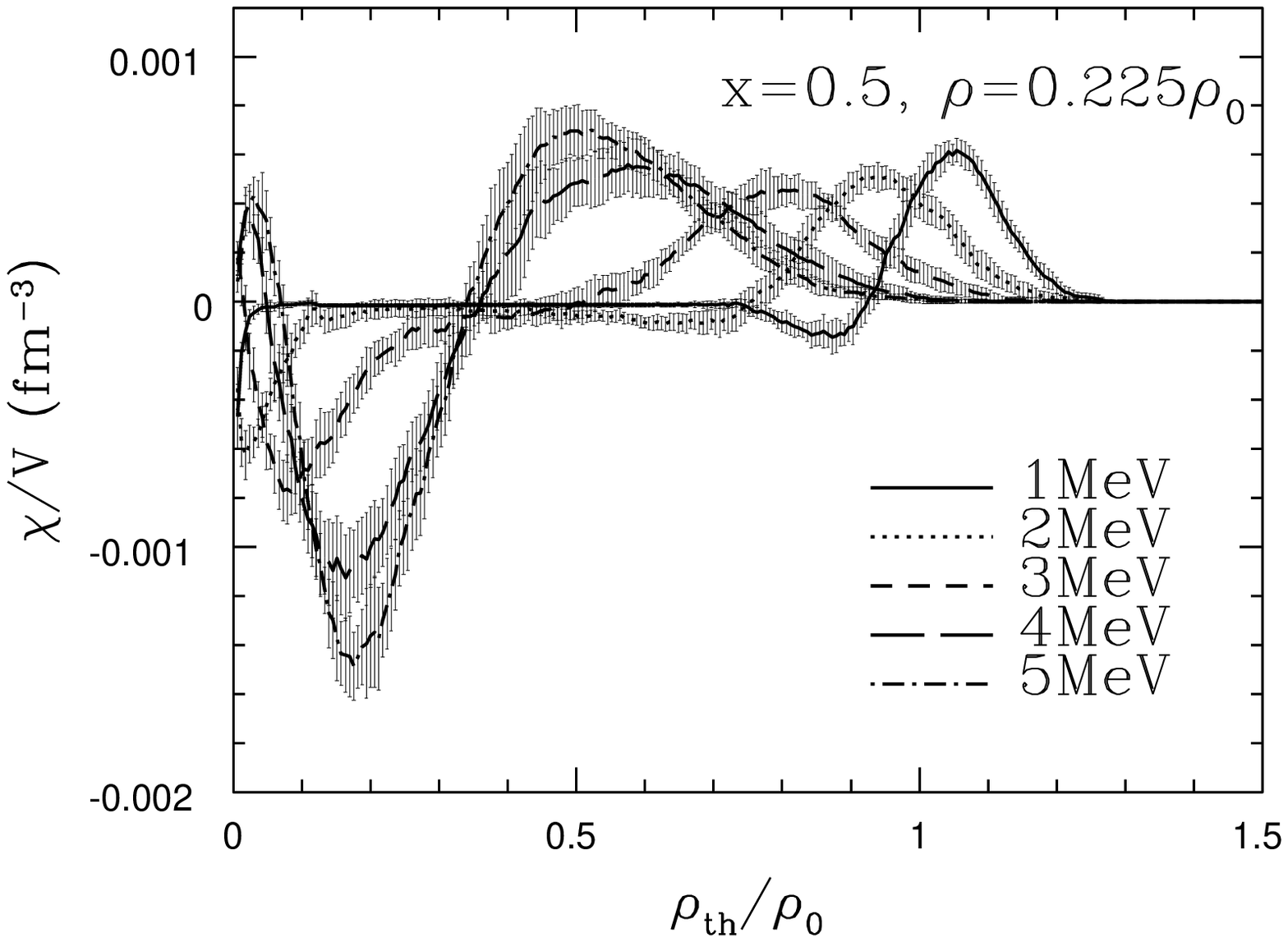}}
\caption{\label{euler 0.225rho x0.5}
  Euler characteristic density for various temperatures plotted 
  as a function of the threshold density $\rho_{\rm th}$.
  The data shown here are calculated for $x=0.5$ and $\rho=0.225\rho_0$,
  where the system is in the phase with rodlike nuclei at zero temperature.
  The error bars are the standard deviations in the long-time average.
  For $T=2$ MeV, the plateau region lies between
  $\rho_{\rm th}\simeq0.45\rho_0$
  and $\simeq0.6\rho_0$, where $\chi/V$ is negative.
  }
\end{center}
\end{figure}
\begin{figure}[htbp]
\begin{center}
\resizebox{8.5cm}{!}
{\includegraphics{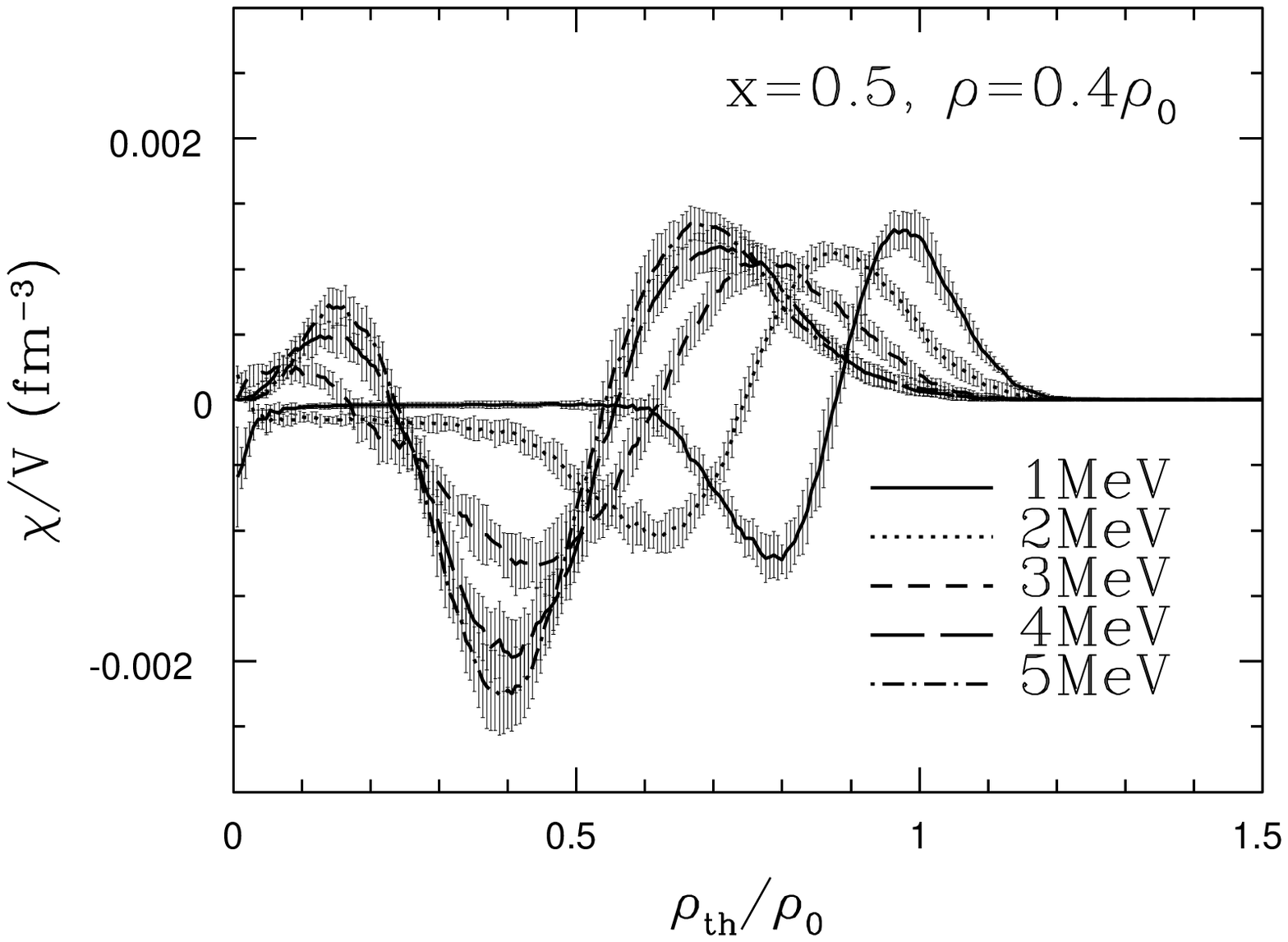}}
\caption{\label{euler 0.4rho x0.5}
  The same as Fig.\ \ref{euler 0.225rho x0.5} for $\rho=0.4\rho_0$,
  where the system is in the phase with slablike nuclei at zero temperature.
  }
\end{center}
\end{figure}
\begin{figure}[htbp]
\begin{center}
\resizebox{8.5cm}{!}
{\includegraphics{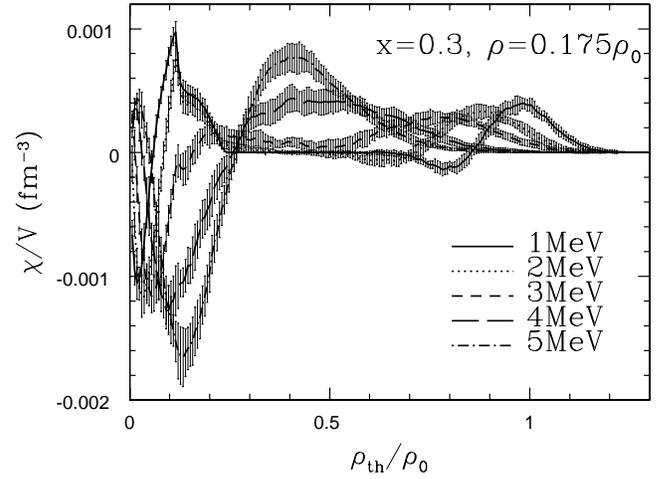}}
\caption{\label{euler 0.175rho x0.3}
  Euler characteristic density for various temperatures plotted 
  as a function of the threshold density $\rho_{\rm th}$.
  The data shown here are calculated for $x=0.3$ and $\rho=0.175\rho_0$,
  where the system is in the phase with rodlike nuclei at zero temperature.
  The error bars are the standard deviations in the long-time average.
  }
\end{center}
\end{figure}
\begin{figure}[htbp]
\begin{center}
\resizebox{8.5cm}{!}
{\includegraphics{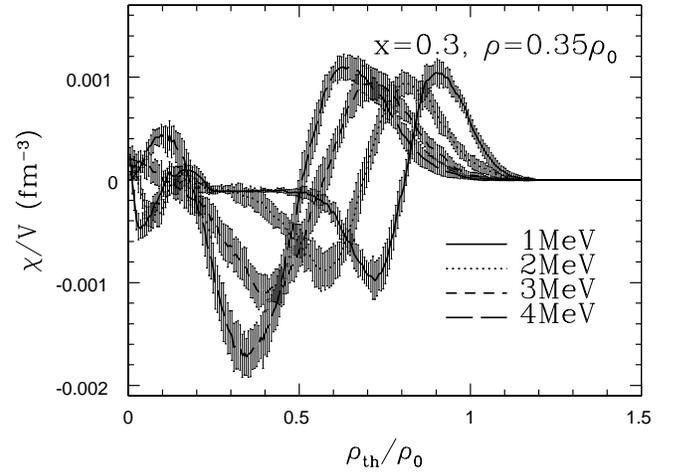}}
\caption{\label{euler 0.35rho x0.3}
  The same as Fig.\ \ref{euler 0.175rho x0.3} for $0.35\rho_0$,
  where the system is in the phase with slablike nuclei at zero temperature.
  }
\end{center}
\end{figure}

To classify the phase in terms of the nuclear structure,
we first have to identify nuclear surface.
The Euler characteristic is useful for identifying the nuclear surface.
Figures \ref{euler 0.225rho x0.5} and \ref{euler 0.4rho x0.5},
show the Euler characteristic density $\chi/V$ of the nucleon density field
calculated for symmetric matter
for $\rho=0.225$ and 0.4$\rho_{0}$, respectively, and they are
plotted as a function of the threshold density $\rho_{\rm th}$
for an isodensity surface.
As for $x=0.3$, $\chi/V$ is plotted in
Figs.\ \ref{euler 0.175rho x0.3} and \ref{euler 0.35rho x0.3}
for $\rho=0.175$ and 0.35$\rho_{0}$, respectively.
It is noted that the curves of $\chi/V$ for lower temperatures
have a plateau, which corresponds to the nuclear surface.
The plateau value of $\chi/V$ is in the range of $-O(10^4)$ to
$O(10^4)$ (see, e.g., Figs.\ 12 and 16 in Ref.\ \cite{qmd2}),
thus we should use the region where the error is smaller than $\sim10^{-4}$
for judging whether $\chi/V$ is zero or not
(hereafter, we use the word ``plateau'' for the plateau region
in which the error is smaller than $\sim10^{-4}$).
According to these figures, the plateau of $\chi/V$ can be observed
at $T=1$ and 2 MeV, while it cannot be seen for $T\ge3$ MeV.
Thus we can say that the nuclear surface cannot be identified
in the temperature range of $T\ge3$ MeV at these values of $x$ and $\rho$
even in the phase-separating region.
This result agrees with the qualitative feature of the snapshots
of nucleon distributions shown
in Figs \ref{snap 0.225rho x0.5 2000} -- \ref{snap 0.4rho x0.5 16000},
\ref{snap 0.175rho x0.3 16000} and \ref{snap 0.34rho x0.3 16000}.

In the region where the nuclear surfaces can be identified,
we sort them into several phases by their nuclear shapes.
We extract morphological features of the nuclear surface
using the Euler characteristic density $\chi/V$
and the averaged mean curvature $\langle H \rangle$.
As one can see from Eq.\ (\ref{euler}),
$\chi/V > 0$ for the sphere and the spherical hole phases,
which have isolated regions and cavities, respectively
(the coexistence phase of spheres and cylinders also has $\chi/V > 0$),
and $\chi/V = 0$ for the other ideal pasta phases
(see the plateau of $\chi/V = 0$ for lower temperatures
shown in Figs.\ \ref{euler 0.225rho x0.5}, \ref{euler 0.4rho x0.5},
\ref{euler 0.175rho x0.3}, and \ref{euler 0.35rho x0.3}),
i.e., the cylinder, the slab and the cylindrical hole phases
which consist of infinitely long rods, infinitely extending slabs and
infinitely long cylindrical holes, respectively.
The ``spongelike'' intermediate phases which are obtained in our previous works
about zero-temperature matter \cite{qmd1,qmd2} have negative values of $\chi/V$
(see Fig.\ 12 of Ref.\ \cite{qmd2}).

As for the averaged mean curvature, at zero temperature,
this quantity decreases almost monotonically from positive to negative
with increasing density until matter becomes uniform
(see Fig.\ 12 of Ref.\ \cite{qmd2}).
Obviously, the zero-value of $\langle H \rangle$ corresponds
to the slab phase.

In our previous work \cite{qmd2},
the sequence of nuclear shapes at zero temperature
with increasing density
is described by the quantities $\langle H \rangle$ and $\chi/V$ as follows:
sphere $(\langle H \rangle > 0,\ \chi/V > 0)$ $\rightarrow$
cylinder $(\langle H \rangle > 0,\ \chi/V = 0)$ $\rightarrow$
$(\langle H \rangle > 0,\ \chi/V < 0)$ $\rightarrow$
slab $(\langle H \rangle = 0,\ \chi/V = 0)$ $\rightarrow$
$(\langle H \rangle < 0,\ \chi/V < 0)$ $\rightarrow$
cylindrical hole $(\langle H \rangle < 0,\ \chi/V = 0)$ $\rightarrow$
spherical hole $(\langle H \rangle < 0,\ \chi/V > 0)$ $\rightarrow$ uniform.

\begin{figure*}[htbp]
\begin{center}
\rotatebox{270}{
\resizebox{9cm}{!}
{\includegraphics{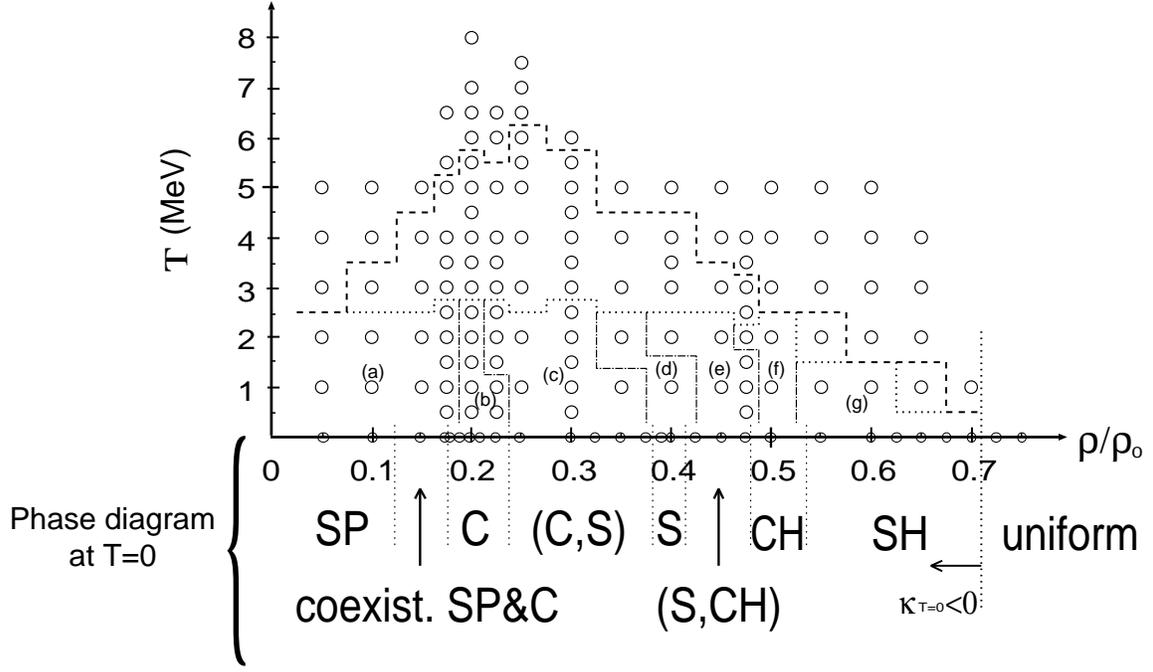}}}
\caption{\label{phase diagram x0.5}
  Phase diagram of matter at $x=0.5$ plotted in the $\rho$ - $T$ plane.
  The dashed and the dotted lines on the diagram
  show the phase separation line and
  the limit below which the nuclear surface can be identified, respectively;
  thus the regions above the dotted line and the dashed line show
  the non-uniform regions, which are difficult to be classified
  into specific phases in terms of the nuclear structure.
  The dash-dotted lines are the phase boundaries between
  the different nuclear shapes.
  The symbols SP, C, S, CH, SH, U stand for nuclear shapes,
  i.e. sphere, cylinder, slab, cylindrical hole,
  spherical hole and uniform, respectively.
  The parentheses (A,B) show intermediate phases between A and B-phases.
  The regions (a)-(g) correspond to the nuclear shapes characterized by
  $\langle H \rangle$ and $\chi/V$ as follows:
  (a) $\langle H \rangle > 0,\ \chi/V > 0$;
  (b) $\langle H \rangle > 0,\ \chi/V = 0$;
  (c) $\langle H \rangle > 0,\ \chi/V < 0$;
  (d) $\langle H \rangle = 0,\ \chi/V = 0$;
  (e) $\langle H \rangle < 0,\ \chi/V < 0$;
  (f) $\langle H \rangle < 0,\ \chi/V = 0$;
  (g) $\langle H \rangle < 0,\ \chi/V > 0$.
  Simulations have been carried out at points denoted by circles.
  }
\end{center}
\end{figure*}

\begin{figure*}[htbp]
\begin{center}
\rotatebox{270}{
\resizebox{9cm}{!}
{\includegraphics{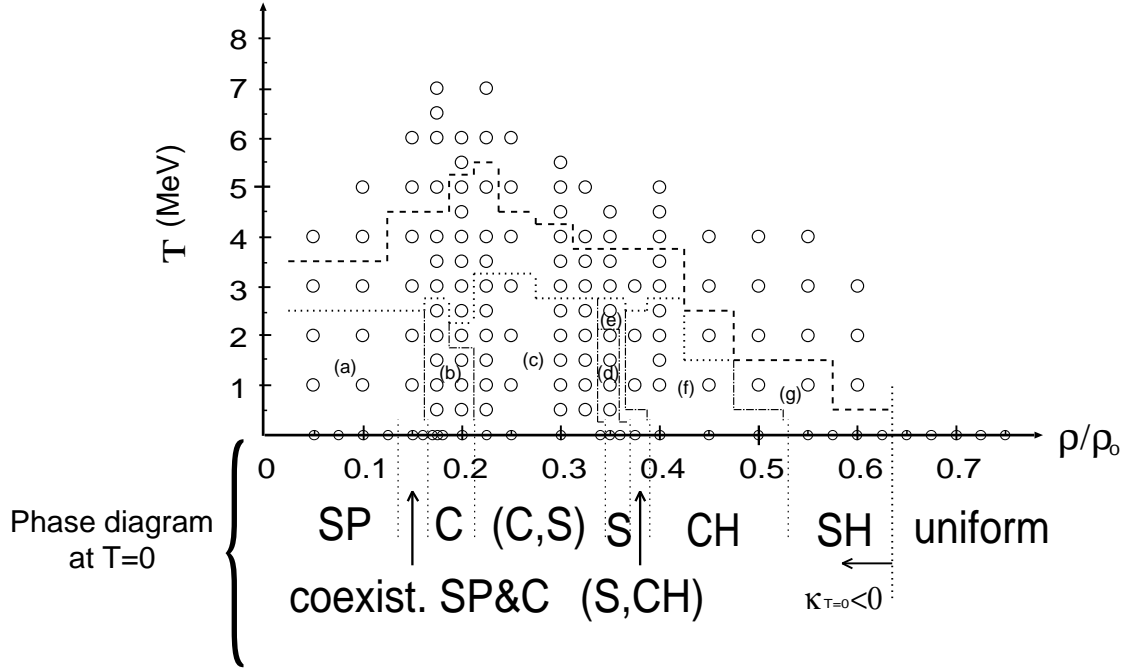}}}
\caption{\label{phase diagram x0.3}
  Phase diagram of matter at $x=0.3$ plotted in the $\rho$ - $T$ plane.
  }
\end{center}
\end{figure*}

Here we classify the nuclear structure at various temperatures and densities
according to the combinations of signs of $\langle H \rangle$ and $\chi/V$
as we have done in the case of $T=0$ \cite{qmd2}.
The phase diagrams obtained for $x=0.5$ and 0.3 are plotted in 
Figs.\ \ref{phase diagram x0.5} and \ref{phase diagram x0.3}, respectively,
in the $\rho$ - $T$ plane \cite{note phase diagram}.
Qualitative features of these phase diagrams are the same,
but the phase separating-region surrounded by a dashed line
is smaller for $x=0.3$ than that for $x=0.5$.
As shown by dotted lines in these figures,
nuclear surface can be identified
typically at $T \alt 3$ MeV in the density range of interest.
In this region where we can observe the nuclear surface,
the nuclear structure generally changes in the same sequence mentioned above
even at finite temperatures
(the only exception is that the density region
of the phase with slablike nuclei is bounded
at $T \agt 2$ MeV for $x=0.3$).
The regions above the dotted line and the dashed line
correspond to some non-uniform phase,
which is however difficult to be classified
into specific phases because the nuclear surface cannot be well identified.

The most striking finite temperature effect on the phase diagrams
is that the density of each phase boundary
between the different structures decreases as $T$ increases.
This is due to the thermal expansion of the nuclear matter region.
At higher temperatures, the volume fraction of the nuclear matter
region is larger even at smaller densities, which leads to
decrease of the each phase boundary
(at $0.225\rho_0$ for $x=0.5$ and at $0.175\rho_0$ for $x=0.3$,
the volume fraction of the nuclear matter region increases
by $\sim 5 \%$ at $T=2$ MeV from the value at $T=1$ MeV;
at $0.4\rho_0$ for $x=0.5$ and at $0.35\rho_0$ for $x=0.3$,
this increase is $\sim 10-15\%$).
This feature is less noticeable for $x=0.3$ than $x=0.5$,
probably because the thermal expansion of nuclei is prevented by
the pressure of the dripped neutrons outside of nuclei.

Above $T \sim 3$ MeV, single-particle excitations are dominant
rather than collective nuclear deformation, and simultaneously,
the binding potential of ``nuclei'' becomes smeared out.
Further increase of temperature leads to dissolution of nuclei
into the uniform fluid phase.
The critical point for the phase separation
is located at $\rho \sim 0.25\rho_0$
and $T \agt 6$ MeV for $x=0.5$ and at $\rho \sim 0.2\rho_0$
and $T \agt 5$ MeV for $x=0.3$.
The above value of the critical temperature for $x=0.5$
is not so different from the result
obtained by Chikazumi {\it et al.}, $T_{c}=8$ MeV,
using a similar nuclear potential \cite{chikazumi}.

As far as we have investigated, the colloidal state in which
nuclei of various sizes and shapes coexist cannot be observed
except for the coexistence phase of spheres and cylinders.
To resolve the problem whether the colloidal phases
are realized due to entropy effect at finite temperatures,
it would be necessary to perform simulations
with the 16384-nucleon system for various densities
other than the typical cases
shown in the Section \ref{snapshots}, or for the temperature region
between 2 and 3 MeV carefully, in which the nuclear surface still remains
but its fluctuations are large.


\section{Summary and Conclusion\label{conclusion}}

We have investigated the structure of hot nuclear matter
at subnuclear densities by QMD simulations.
The values of the proton fraction studied in the present work are 0.5 and 0.3;
the latter is a typical value in supernova inner cores.
The phase diagram obtained for $x=0.3$ (Fig.\ \ref{phase diagram x0.3})
tells that the pasta phases and the spongelike intermediate phases
survive until $T \alt 3$ MeV except for the higher density region
of $\rho \agt 0.45\rho_{0}$.
This result strongly suggests that these phases with exotic nuclear structures
can exist in the inner cores in the pre-bounce stage of
the collapse-driven supernovae.
Thus it would be meaningful to investigate a modification
to the neutrino opacity due to the infinitely extending ``pasta''
and spongelike structures in the future \cite{opacity}.

The present research has provided the general picture of
the phase diagram at subnuclear densities in the temperature vs density plane.
The qualitative features of the nucleon distribution at finite temperatures,
except for higher densities just below which matter becomes uniform
even at $T \sim 0$ MeV,
may be summarized as follows.
\begin{itemize}
\item
At $T\simeq$ 1--1.5 MeV, the number of evaporated nucleons 
starts to be significant. However, the macroscopic structure of matter
does not basically change from that of zero-temperature matter
except for a small expansion of the nuclear matter regions.
\item
At $T\simeq$ 1.5--2.5 MeV,
fluctuations of the nuclear shape start to be significant
and, in some cases, the topology of the nuclear structure changes
from that in the zero-temperature matter for the same density.
\item
At $T\simeq$ 2.5 -- 3 MeV, single particle excitations are
dominant rather than collective excitations, representing nuclear deformations.
The density profile of nucleons is smoothed out,
and nuclear surface cannot be identified by a threshold density.
At higher temperatures, even the density contrast becomes drastically smaller,
which leads to the transition to uniform matter.
The order of the transition from inhomogeneous matter to uniform matter
has yet to be clarified.
\item
The critical temperature $T_{\rm c}$ for the phase separation
is $\agt 6$ MeV for $x=0.5$ and $\agt 5$ MeV for $x=0.3$.
\end{itemize}

Through the present work and the previous ones \cite{qmd1,qmd2},
we have depicted the structure of nuclear matter
in neutron star inner crusts and supernova inner cores
in its equilibrium state.
Our studies will be helpful in understanding the realistic situations
of the interior of the dense stars.
For further work, studies on the dynamical aspects of the nuclear pasta
become important.
For example, in connection with the possible instability of
the long-range order of the planar phase, which cannot be
fully incorporated in our simulations,
elastic properties \cite{pp} of the pasta and spongelike phases
should be studied further.
Structural transitions between the pasta phases
induced by compression and decompression is one of the most interesting
problems \cite{formation} which have not yet been studied at all.
We believe that our work here,
which was performed within a dynamical framework,
will open up these new aspects of the study on nuclear pasta.

\begin{acknowledgments}
G. W. is grateful to T. Maruyama, K. Iida and I. Kayo
for helpful discussions and comments.
G. W. also appreciates C. J. Pethick for his valuable comments and
hospitality at NORDITA and
M. Shimizu, L. M. Jensen, and P. Urkedal
for arranging the computer environment.
This work was supported in part
by the Junior Research Associate Program in RIKEN
through Research Grant No. J130026,
by Grants-in-Aid for Scientific Research
provided by the Ministry of
Education, Culture, Sports, Science and Technology
through Research Grant (S) No. 14102004, No. 14079202 and No. 14-7939
and by Nishina Memorial Foundatoin.
\end{acknowledgments}


\begin{thebibliography}{99}
\bibitem{note vpe} Though the term $V_{p-e}$ in $V_{\rm Coulomb}$
is not explicitly written in Eq.\ (12) of Ref.\ \cite{qmd2},
this contribution is included in the simulations in Refs.\ \cite{qmd1,qmd2}
by the Ewald summation (see Appendix A of Ref.\ \cite{qmd2}).
\bibitem{note fluctuation} In the present simulations,
thermal fluctuations with a wavelength longer than the box size
cannot be incorporated. We should remark that
these modes could be effective at destroying
the long-range order of the phase with slablike nuclei.
\end{thebibliography}

\end{document}